\documentclass [11pt,a4paper] {article}

\usepackage[cp1252]{inputenc}

\usepackage{amssymb}
\usepackage{amsmath}
\usepackage{amsfonts,amssymb}

\usepackage[dvips]{graphicx}

\usepackage{bbm}

\DeclareMathAlphabet{\mathpzc}{OT1}{pzc}{m}{it}

\begin{document}

\title{Quantum supervaluationist account of the EPR paradox}

\author{Arkady Bolotin\footnote{$Email: arkadyv@bgu.ac.il$\vspace{5pt}} \\ \textit{Ben-Gurion University of the Negev, Beersheba (Israel)}}

\maketitle

\begin{abstract}\noindent In the paper, the EPR paradox is explored by the approach of quantum supervaluationism that leads to a ``gappy'' semantics with the propositions giving rise to truth-value gaps. Within this approach, the statement, which asserts that in the singlet state the system of two (i.e., $A$ and $B$) spin-$\textonehalf$ particles possesses the a priori property ``spin $A$ is up and spin $B$ is down along the same axis'' or ``spin $A$ is down and spin $B$ is up along the same axis'', does not have the truth-value at all. Consequently, after the verification of, say, the proposition ``spin $A$ is up along the $z$-axis'', the statistical population describing the valuation of the logical connective ``spin $B$ is down along the $z$-axis and spin $B$ is up (down) along the $x$-axis'' would have no elements.\\

\noindent \textbf{Keywords:} Quantum mechanics; EPR paradox; Truth values; Bivalence; Supervaluationism.\\

\end{abstract}

\section{Introduction}  

\noindent Let $|s,m_j\rangle$ denote the vector of the Hilbert space describing the state of particle's spin, where $s$ stands for the spin quantum number and $m_j$ specifies the spin projection quantum number along the $j \in\{x,y,z\}$ axis. Consider a system with two ($A$ and $B$) spin-$\textonehalf$ particles (that is, $s^{(A)}=s^{(B)}=\textonehalf$) which is prepared in a singlet state (i.e., a state with total spin angular momentum 0) described by the vector $|0,0_j\rangle$, namely,\smallskip

\begin{equation} \label{1} 
   |0,0_j\rangle
   =
   \frac{1}{\sqrt{2}}
   \left(
      \Big|\textonehalf,m^{(A)}_j\!\!=+\textonehalf\Big\rangle \!\otimes\! \Big|\textonehalf,m^{(B)}_j\!\!=-\textonehalf\Big\rangle
      -
      \Big|\textonehalf,m^{(A)}_j\!\!=-\textonehalf\Big\rangle \!\otimes\! \Big|\textonehalf,m^{(B)}_j\!\!=+\textonehalf\Big\rangle      
   \right)
   \;\;\;\;  .
\end{equation}
\smallskip

\noindent Suppose that after being prepared in the singlet, these particles travel away from each other in a region of zero magnetic field where by means of a Stern-Gerlach magnet an observer ``Alice'' measures the spin of the particle $A$ and by means of another Stern-Gerlach magnet an observer ``Bob'' measures the spin of the particle $B$. Assume that the measurements are space-like separated, such that neither of the observers can act upon or exercise influence on the result of the other.\\

\noindent Let $\uparrow_j$ denote the proposition asserting that the spin-$\textonehalf$ particle exists in the spin state ``$j$-up'' (i.e., the statement ``$m_j =+\textonehalf$'') whereas $\downarrow_j$ signify the alternative proposition that this particle is in the spin state ``$j$-down'' (i.e., the statement  ``$m_j =-\textonehalf$'').\\

\noindent Let the double-bracket notation ${[\![ \,\diamond\, ]\!]}_v$ where the symbol $\diamond$ stands for any proposition (compound or simple) denote \textit{a valuation in a circumstance} $v$, that is, a mapping from a set of propositions $\{\diamond\}$ to a set of truth-values $\mathcal{V}_N = \{\mathfrak{v}\}$ having the cardinality $N$ and the range with the upper bound 1 (which represents \textit{the truth}) and the lower bound 0 (representing \textit{the falsehood}), relative to a particular circumstance of evaluation indicated by $v$.\\

\noindent Let us consider the propositions $\mathrm{Same}_j$ and $\mathrm{Diff}_j$:\smallskip

\begin{equation} \label{2} 
   {[\![ \, \mathrm{Same}_j \, ]\!]}_v
   =
   {[\![ \,
      \uparrow^{(A)}_j \!\!\!\wedge\! \uparrow^{(B)}_j
      \underline{\lor}\,\,
      \downarrow^{(A)}_j\! \!\!\wedge\! \downarrow^{(B)}_j
   ]\!]}_v
   \;\;\;\;  ,
\end{equation}

\begin{equation} \label{3} 
   {[\![ \, \mathrm{Diff}_j \, ]\!]}_v
   =
   {[\![ \,
      \uparrow^{(A)}_j \!\!\!\wedge\! \downarrow^{(B)}_j
      \underline{\lor}\,\,
      \downarrow^{(A)}_j\! \!\!\wedge\! \uparrow^{(B)}_j
   ]\!]}_v
   \;\;\;\;  ,
\end{equation}
\smallskip

\noindent where $\underline{\lor}$ stands for ``\textit{exclusive or}'' logical connective. The proposition $\mathrm{Same}_j$ asserts that the two particles have the same directions of their spins along the $j$-axis, while the proposition $\mathrm{Diff}_j$ declares that their spin directions are different along $j$.\\

\noindent Because the system is prepared in the singlet state, Alice and Bob can affirm that prior to the verification the proposition $\mathrm{Same}_j$ has the value of falsehood at the same time as the proposition $\mathrm{Diff}_j$ has the value of truth, i.e., ${[\![ \, \mathrm{Same}_j \, ]\!]}_v = 0$ and ${[\![ \, \mathrm{Diff}_j \, ]\!]}_v = 1$.\\

\noindent Suppose that using the outcome of the measurement, Alice proves (disproves) the statement ``$m^{(A)}_j\!\!=+\textonehalf$'' as well as the statement ``$m^{(A)}_j\!\!=-\textonehalf$''. This act destroys information about the projection of the spin of the particle $A$ along any other axis $k\in\{x,y,z\}$ not equal to $j$ (that might previously have been obtained) and, for that reason, one can write\smallskip

\begin{equation} \label{4} 
   \left\{
   {[\![ \, \updownarrow^{(A)}_j  ]\!]}_v
   \right\}
   =
   \{0,1\}
   \;\;
   \implies
   \;\;
   \left\{
   {[\![ \, \updownarrow^{(A)}_{k \neq j}  ]\!]}_v
   \right\}
   \neq
   \{0,1\}
   \;\;\;\;  ,
\end{equation}
\smallskip

\noindent where the symbol $\updownarrow$ must be replaced by either $\uparrow$ or $\downarrow$.\\

\noindent However, if both of the following premises are assumed, namely,

\begin{description}  
\item \quad (\textbf{1}) \textit{the truth-values of the propositions $\updownarrow_j$ exist before the act of verification},
\item \quad (\textbf{2}) \textit{these truth-values are elements of the two-valued set $\mathcal{V}_2 = \{0,1\}$},
\end{description}

\noindent then the valuation ${[\![ \, \mathrm{Diff}_j \, ]\!]}_v=1$ will imply\smallskip

\begin{equation} \label{5} 
   {[\![ \, \uparrow^{(A)}_j  ]\!]}_v \!\cdot {[\![ \, \downarrow^{(B)}_j  ]\!]}_v
   +
   {[\![ \, \downarrow^{(A)}_j  ]\!]}_v \!\cdot {[\![ \, \uparrow^{(B)}_j  ]\!]}_v
   =
   1
   \;\;\;\;  ,
\end{equation}
\smallskip

\noindent or, explicitly,\smallskip

\begin{equation} \label{6} 
   {[\![ \, \downarrow^{(B)}_j  ]\!]}_v
   =
   {[\![ \, \uparrow^{(A)}_j  ]\!]}_v
   \;\;\;\;  ,
\end{equation}

\begin{equation} \label{7} 
   {[\![ \, \uparrow^{(B)}_j  ]\!]}_v
   =
   {[\![ \, \downarrow^{(A)}_j  ]\!]}_v
   \;\;\;\;  .
\end{equation}
\smallskip

\noindent Hence, the verification of $\updownarrow^{(A)}_j$ will produce the bivaluation $\{  {[\![ \, \updownarrow^{(B)}_j  ]\!]}_v \}=\{0,1\}$ without destroying information about the spin projection of the particle $B$ along any axis $k \neq j$.\\

\noindent Let us consider, for example, the product ${[\![ \, \downarrow^{(B)}_z  ]\!]}_v \!\cdot {[\![ \, \uparrow^{(B)}_x  ]\!]}_v$. Its statistical population is the ``cross product'' of the sets  $\{{[\![ \, \downarrow^{(B)}_z  ]\!]}_v\} \!\times\! \{{[\![ \, \uparrow^{(B)}_x  ]\!]}_v\}$ and defined such that\smallskip

\begin{equation} \label{8} 
   \{{[\![ \, \downarrow^{(B)}_z  ]\!]}_v\} \!\times\! \{{[\![ \, \uparrow^{(B)}_x  ]\!]}_v\}
   =
   \left\{
      \left(
         \{{[\![ \, \downarrow^{(B)}_z  ]\!]}_v\} , \{{[\![ \, \uparrow^{(B)}_x  ]\!]}_v\}
      \right)
   \right\}
   \;\;\;\;  .
\end{equation}
\smallskip

\noindent Following the verification of the proposition $\uparrow^{(A)}_z$, the said statistical population will contain two pairs -- each for every possible preexisitng truth-value of $\uparrow^{(B)}_x$, namely,\smallskip

\begin{equation} \label{9} 
   \{{[\![ \, \downarrow^{(B)}_z  ]\!]}_v\} \!\times\! \{{[\![ \, \uparrow^{(B)}_x  ]\!]}_v\}
   =
   \{1\} \!\times\! \{{[\![ \, \uparrow^{(B)}_x  ]\!]}_v\}
   =
   \left\{
      \left(
         1 , 1
      \right)
      ,
      \left(
         1 , 0
      \right)
   \right\}
   \;\;\;\;  .
\end{equation}
\smallskip

\noindent In consequence, the verification (refutation) of $\uparrow^{(A)}_x$ carried out in another experimental run will bring the definiteness to this product. This way, the statement ``$m^{(B)}_z\!\!=-\textonehalf$ and $m^{(B)}_x\!\!=+\textonehalf$'' will have the truth value contrary to the basic principles of quantum theory.\\

\noindent As it can be readily seen, to resolve this paradox (known as \textit{the EPR paradox} \cite{Einstein}) upon the supposition that the particle $B$ cannot be affected by measurements carried out on the particle $A$, one can deny either the premise (\textbf{1}) or the premise (\textbf{2}).\\

\noindent This paper presents a logic approach to the EPR paradox where the premise (\textbf{1}) is denied.\\

\section{Quantum supervaluationism}  

\noindent Consider the lattice $L(\mathcal{H})$ formed by the column spaces (ranges) of the projection operators $\hat{P}_{\alpha}$, $\hat{P}_{\beta}$, $\dots$ on the Hilbert space $\mathcal{H}$.\\ 

\noindent In the lattice $L(\mathcal{H})$ the ordering relation $\le$ corresponds to the subset relation $\mathrm{ran}(\hat{P}_{\alpha}) \subseteq \mathrm{ran}(\hat{P}_{\beta})$; the operation \textit{meet} $\sqcap$ corresponds to the interception $\mathrm{ran}(\hat{P}_{\alpha}) \cap \mathrm{ran}(\hat{P}_{\beta})$; the operation \textit{join} $\sqcup$ corresponds to the smallest closed subspace of $\mathcal{H}$ containing the union $\mathrm{ran}(\hat{P}_{\alpha}) \cup \mathrm{ran}(\hat{P}_{\beta})$. The lattice $L(\mathcal{H})$ is bounded, i.e., it has the greatest element $\mathrm{ran}(\hat{1})=\mathcal{H}$ and the least element $\mathrm{ran}(\hat{0})=\{0\}$ that satisfy the following subset relation for every $\mathrm{ran}(\hat{P})$ in $L(\mathcal{H})$:\smallskip

\begin{equation} \label{10} 
   \mathrm{ran}(\hat{0})
   \subseteq
   \mathrm{ran}(\hat{P})
   \subseteq
   \mathrm{ran}(\hat{1})
   \;\;\;\;  .
\end{equation}
\smallskip

\noindent Let $\hat{P}_{\alpha} \sqcap \hat{P}_{\beta}$, $\hat{P}_{\alpha} \sqcup \hat{P}_{\beta}$ and $\hat{P}_{\alpha} + \hat{P}_{\beta}$ denote projections on the interception $\mathrm{ran}(\hat{P}_{\alpha}) \cap \mathrm{ran}(\hat{P}_{\beta})$, the union $\mathrm{ran}(\hat{P}_{\alpha}) \cup \mathrm{ran}(\hat{P}_{\beta})$ and the sum $\mathrm{ran}(\hat{P}_{\alpha}) + \mathrm{ran}(\hat{P}_{\beta})$, respectively. One can write then\smallskip

\begin{equation} \label{11} 
   \mathrm{ran}
      (
         \hat{P}_{\alpha} \sqcap \hat{P}_{\beta}
      )
   =
   \mathrm{ran}(\hat{P}_{\alpha})
   \cap
   \mathrm{ran}(\hat{P}_{\beta})
   \;\;\;\;  ,
\end{equation}

\begin{equation} \label{12} 
   \mathrm{ran}
      (
         \hat{P}_{\alpha} \sqcup \hat{P}_{\beta}
      )
   =
   \mathrm{ran}(\hat{P}_{\alpha})
   \cup
   \mathrm{ran}(\hat{P}_{\beta})
   \;\;\;\;  ,
\end{equation}

\begin{equation} \label{13} 
   \mathrm{ran}
      (
         \hat{P}_{\alpha} + \hat{P}_{\beta}
      )
   =
   \mathrm{ran}(\hat{P}_{\alpha})
   +
   \mathrm{ran}(\hat{P}_{\beta})
   \;\;\;\;  .
\end{equation}
\smallskip

\noindent Clearly, if the column spaces $\mathrm{ran}(\hat{P}_{\alpha})$ and $\mathrm{ran}(\hat{P}_{\beta})$ are orthogonal, their union coincides with their sum, i.e.,

\begin{equation} \label{14} 
   \mathrm{ran}(\hat{P}_{\alpha})
   \cap
   \mathrm{ran}(\hat{P}_{\beta})
   =
   \mathrm{ran}(\hat{0})
   \;\;
   \implies
   \;\;
   \mathrm{ran}(\hat{P}_{\alpha})
   \cup
   \mathrm{ran}(\hat{P}_{\beta})
   =
   \mathrm{ran}(\hat{P}_{\alpha})
   +
   \mathrm{ran}(\hat{P}_{\beta})
   \;\;\;\;  .
\end{equation}
\smallskip

\noindent On the other hand, if the projection operators $\hat{P}_{\alpha}$ and $\hat{P}_{\beta}$ on $\mathcal{H}$ are orthogonal, then $\hat{P}_{\alpha}\hat{P}_{\beta}=\hat{P}_{\beta}\hat{P}_{\alpha}=\hat{0}$. Hence, in this case one must get\smallskip

\begin{equation} \label{15} 
   \hat{P}_{\alpha} \sqcap \hat{P}_{\beta}
   =
   \hat{P}_{\alpha} \hat{P}_{\beta}
   =
   \hat{0}
   \;\;\;\;  ,
\end{equation}

\begin{equation} \label{16} 
   \hat{P}_{\alpha} \sqcup \hat{P}_{\beta}
   =
   \hat{P}_{\alpha} + \hat{P}_{\beta}
   \;\;\;\;  .
\end{equation}
\smallskip

\noindent Now, consider the truth-value assignments of the projection operators in the lattice $L(\mathcal{H})$.\\

\noindent Let $v$ be the truth-value assignment function and $\hat{P}_{\diamond}$ denote the projection operator associated with the proposition $\diamond$. Assume that the following valuational axiom holds:\smallskip

\begin{equation} \label{17} 
   v(\hat{P}_{\diamond})
   =
   {[\![ \,\diamond\, ]\!]}_v
   \;\;\;\;  .
\end{equation}
\smallskip

\noindent Suppose that a system is in a pure state $|{\Psi}_{\alpha}\rangle$ lying in the column space of the projection operator $\hat{P}_{\alpha}$. Since being in $\mathrm{ran}(\hat{P}_{\alpha})$ means $\hat{P}_{\alpha}|{\Psi}_{\alpha}\rangle = 1 \cdot |{\Psi}_{\alpha}\rangle$, one can assume that in the state $|{\Psi}_{\alpha}\rangle \in \mathrm{ran}(\hat{P}_{\alpha})$, the truth-value assignment function $v$ assigns the truth value 1 to the operator $\hat{P}_{\alpha}$ and, in this way, the proposition $\alpha$, specifically, $v(\hat{P}_{\alpha}) = {[\![ \,\alpha\, ]\!]}_v = 1$. Contrariwise, if $v(\hat{P}_{\alpha}) = {[\![ \,\alpha\, ]\!]}_v = 1$, then one can assume that the system is in the state $|{\Psi}_{\alpha}\rangle \in \mathrm{ran}(\hat{P}_{\alpha})$. These two assumptions can be written down together as the logical biconditional, namely,\smallskip

\begin{equation} \label{18} 
   |{\Psi}_{\alpha}\rangle \in \mathrm{ran}(\hat{P}_{\alpha})
   \;\;
   \iff
   \;\;
   v(\hat{P}_{\alpha})
   =
   {[\![ \,\alpha\, ]\!]}_v
   =
   1
   \;\;\;\;  .
\end{equation}
\smallskip

\noindent On the other hand, the vector $|{\Psi}_{\alpha}\rangle$ must lie in the null space of any projection operator $\hat{P}_{\beta}$ orthogonal to $\hat{P}_{\alpha}$. Since being in $\mathrm{ker}(\hat{P}_{\beta})$ means $\hat{P}_{\beta}|{\Psi}_{\alpha}\rangle = 0 \cdot |{\Psi}_{\alpha}\rangle$, one can assume then that\smallskip

\begin{equation} \label{19} 
   |{\Psi}_{\alpha}\rangle \in \mathrm{ker}(\hat{P}_{\beta})
   \;\;
   \iff
   \;\;
   v(\hat{P}_{\beta})
   =
   {[\![ \,\beta\, ]\!]}_v
   =
   0
   \;\;\;\;  .
\end{equation}
\smallskip

\noindent Suppose by contrast that the system is in the pure state $|{\Psi}\rangle$ that does not lie in the column or null space of the projection operator $\hat{P}_{\diamond}$, i.e., $|{\Psi}\rangle \notin \mathrm{ran}(\hat{P}_{\diamond})$ and $|{\Psi}\rangle \notin \mathrm{ker}(\hat{P}_{\diamond})$. Under the valuation assumptions (\ref{18}) and (\ref{19}), the truth-value function $v$ must assign neither 1 nor 0 to $\hat{P}_{\diamond}$, namely, $v(\hat{P}_{\diamond}) \neq 1$ and $v(\hat{P}_{\diamond}) \neq 0$. Hence, in this case the proposition $\diamond$ associated with $\hat{P}_{\diamond}$ cannot be bivalent, namely, ${[\![ \,\diamond\, ]\!]}_v \notin \mathcal{V}_2$.\\

\noindent Using a supervaluationary semantics (see, for example \cite{Varzi} or \cite{Keefe}), this failure of bivalence can be described as \textit{a truth-value gap} for the proposition $\diamond$, explicitly,\smallskip

\begin{equation} \label{20} 
   |\Psi\rangle
   \notin
   \left\{
      \begin{array}{l}
         \mathrm{ran}(\hat{P}_{\diamond})\\
         \mathrm{ker}(\hat{P}_{\diamond})
      \end{array}
   \right.
   \iff
   \;
   \{v(\hat{P}_{\diamond})\}
   =
   \{{[\![ \,\diamond\, ]\!]}_v\}
   =
   \varnothing
   \;\;\;\;  .
\end{equation}
\smallskip

\noindent Within the said semantics, the operators $\hat{1}$ and $\hat{0}$ can be equated with ``\textit{the super-truth}'' and ``\textit{the super-falsity}'' since under the valuations (\ref{18}) and (\ref{19}) these operators are true and false, respectively, in any arbitrary state $|\Phi\rangle$ in the Hilbert space $\mathcal{H}$, that is,\smallskip

\begin{equation} \label{21} 
   |\Phi\rangle
   \in
   \left\{
      \begin{array}{l}
         \mathrm{ran}(\hat{1}) = \mathcal{H}\\
         \mathrm{ker}(\hat{0}) = \mathcal{H}
      \end{array}
   \right.
   \iff
   \left\{
      \begin{array}{l}
         v(\hat{1}) = 1\\
         v(\hat{0}) = 0
      \end{array}
   \right.
   \;\;\;\;  .
\end{equation}
\smallskip

\noindent In this way, the approach based on the assumption (\ref{20}) results in a ``gappy'' logic with the propositions giving rise to truth-value gaps. Accordingly, one can call this approach \textit{quantum supervaluationism} (for other details of the approach see \cite{Bolotin}).\\

\section{Logic EPR account}  

\noindent The entangled particles $A$ and $B$ can be represented by observables ${\sigma}_{j}^{(A)}\!\otimes {\sigma}_{j}^{(B)}$ created by the Pauli matrices, namely,\smallskip

\begin{equation} \label{22} 
   {\sigma}_{z}^{(A)}\!\otimes {\sigma}_{z}^{(B)}\!
   =
   \!\left[
      \!\!\!
      \begin{array}{c c c c}
         1 &          0 &          0 & 0\\
         0 & \bar{1} &          0 & 0\\
         0 &          0 & \bar{1} & 0\\
         0 &          0 &          0 & 1
      \end{array}
      \!\!\!
   \right]
   ,
   \;
   {\sigma}_{x}^{(A)}\!\otimes {\sigma}_{x}^{(B)}\!
   =
   \!\left[
      \!\!\!
      \begin{array}{c c c c}
         1 &          0 &          0 & 1\\
         0 &          0 &          1 & 0\\
         0 &          1 &          0 & 0\\
         1 &          0 &          0 & 1
      \end{array}
      \!\!\!
   \right]
   ,
   \;
   {\sigma}_{y}^{(A)}\!\otimes {\sigma}_{y}^{(B)}\!
   =
   \!\left[
      \!\!\!
      \begin{array}{c c c c}
                  1 &          0 &          0 & \bar{1}\\
                  0 &          0 &          1 &          0\\
                  0 &          1 &          0 &          0\\
         \bar{1} &          0 &          0 &          0
      \end{array}
      \!\!\!
   \right]
   \;\;\;\;  ,
\end{equation}
\smallskip

\noindent where $\bar{1} \equiv -1$.\\ 

\noindent Take the observable ${\sigma}_{z}^{(A)}\!\otimes {\sigma}_{z}^{(B)}$: Its eigenvectors $|\Psi_{zz}(\lambda_{1,2})\rangle$ corresponding to the negative eigenvalues $\lambda_{1,2} =-1$ are\smallskip

\begin{equation} \label{23} 
   |\Psi_{zz}(\lambda_{1}=-1)\rangle
   =
   \left[
      \!\!\!
      \begin{array}{l}
         0\\
         1\\
         0\\
         0
      \end{array}
      \!\!\!
   \right]
   \in
   \mathrm{ran}(\hat{P}_{z}^{\uparrow\downarrow})
   =
   \left\{
   \left[
      \!\!\!
      \begin{array}{r}
                0\\
                a\\
                0\\
                0
      \end{array}
      \!\!\!
   \right]\!
   :\,
   a \in \mathbb{R}
   \right\}
   \;\;\;\;  ,
\end{equation}

\begin{equation} \label{24} 
   |\Psi_{zz}(\lambda_{2}=-1)\rangle
   =
   \left[
      \!\!\!
      \begin{array}{l}
         0\\
         0\\
         1\\
         0
      \end{array}
      \!\!\!
   \right]
   \in
   \mathrm{ran}(\hat{P}_{z}^{\downarrow\uparrow})
   =
   \left\{
   \left[
      \!\!\!
      \begin{array}{r}
                0\\
                0\\
                a\\
                0
      \end{array}
      \!\!\!
   \right]\!
   :\,
   a \in \mathbb{R}
   \right\}
   \;\;\;\;  ,
\end{equation}
\smallskip

\noindent where $\hat{P}_{z}^{\uparrow\downarrow}$ and $\hat{P}_{z}^{\downarrow\uparrow}$ denote the orthogonal projection operators defined as\smallskip

\begin{equation} \label{25} 
   \hat{P}_{z}^{\uparrow\downarrow}
   =
   |\!\uparrow_{z}^{(A)}\rangle \langle\uparrow_{z}^{(A)}\!\!|
   \otimes
   |\!\downarrow_{z}^{(B)}\rangle \langle\downarrow_{z}^{(B)}\!\!|
   =
   \left[
      \!\!\!
      \begin{array}{c c c c}
         0 & 0 & 0 & 0\\
         0 & 1 & 0 & 0\\
         0 & 0 & 0 & 0\\
         0 & 0 & 0 & 0
      \end{array}
      \!\!\!
   \right]
   \;\;\;\;  ,
\end{equation}

\begin{equation} \label{26} 
   \hat{P}_{z}^{\downarrow\uparrow}
   =
   |\!\downarrow_{z}^{(A)}\rangle \langle\downarrow_{z}^{(A)}\!\!|
   \otimes
   |\!\uparrow_{z}^{(B)}\rangle \langle\uparrow_{z}^{(B)}\!\!|
   =
   \left[
      \!\!\!
      \begin{array}{c c c c}
         0 & 0 & 0 & 0\\
         0 & 0 & 0 & 0\\
         0 & 0 & 1 & 0\\
         0 & 0 & 0 & 0
      \end{array}
      \!\!\!
   \right]
   \;\;\;\;  .
\end{equation}
\smallskip

\noindent On the other hand, the singlet state $|0,0_z\rangle$ is the vector\smallskip

\begin{equation} \label{27} 
   |0,0_z\rangle
   =
   \frac{1}{\sqrt{2}}
   \left(
      |\!\uparrow_{z}^{(A)}\rangle \otimes |\!\downarrow_{z}^{(B)}\rangle
      -
      |\!\downarrow_{z}^{(A)}\rangle \otimes |\!\uparrow_{z}^{(B)}\rangle     
   \right)
   \;\;\;\;  ,
\end{equation}
\smallskip

\noindent where\smallskip

\begin{equation} \label{28} 
   |\!\uparrow_{z}^{(A)}\rangle \otimes |\!\downarrow_{z}^{(B)}\rangle
   =
   \left[
      \!\!\!
      \begin{array}{l}
         0\\
         1\\
         0\\
         0
      \end{array}
      \!\!\!
   \right]
   \;\;\;\;  ,
\end{equation}

\begin{equation} \label{29} 
   |\!\downarrow_{z}^{(A)}\rangle \otimes |\!\uparrow_{z}^{(B)}\rangle
   =
   \left[
      \!\!\!
      \begin{array}{l}
         0\\
         0\\
         1\\
         0
      \end{array}
      \!\!\!
   \right]
   \;\;\;\;  ,
\end{equation}
\smallskip

\noindent and so\smallskip

\begin{equation} \label{30} 
   |0,0_z\rangle
   =
   \frac{1}{\sqrt{2}}\!\!
   \left(
      \left[
         \!\!\!
         \begin{array}{l}
            0\\
            1\\
            0\\
            0
         \end{array}
         \!\!\!
      \right]
      -
      \left[
         \!\!\!
         \begin{array}{l}
            0\\
            0\\
            1\\
            0
         \end{array}
         \!\!\!
      \right]
   \right)
   =
   \frac{1}{\sqrt{2}}\!\!
      \left[
         \!\!\!
         \begin{array}{l}
                     0\\
                     1\\
            \bar{1}\\
                     0
         \end{array}
         \!\!\!
      \right]
   \;\;\;\;  .
\end{equation}
\smallskip

\noindent As follows, the vector $|0,0_z\rangle$ lies in the column space of the sum of two projection operators $\hat{P}_{z}^{\uparrow\downarrow}$ and $\hat{P}_{z}^{\downarrow\uparrow}$:\smallskip

\begin{equation} \label{31} 
   |0,0_z\rangle
   =
   \frac{1}{\sqrt{2}}\!\!
      \left[
         \!\!\!
         \begin{array}{l}
                     0\\
                     1\\
            \bar{1}\\
                     0
         \end{array}
         \!\!\!
      \right]
   \in
   \mathrm{ran}(\hat{P}_{z}^{\uparrow\downarrow} + \hat{P}_{z}^{\downarrow\uparrow})
   =
   \mathrm{ran}\!\!
   \left(
      \left[
         \!\!\!
         \begin{array}{c c c c}
            0 & 0 & 0 & 0\\
            0 & 1 & 0 & 0\\
            0 & 0 & 1 & 0\\
            0 & 0 & 0 & 0
         \end{array}
         \!\!\!
      \right]
   \right)
   =
   \left\{
   \left[
      \!\!\!
      \begin{array}{r}
                0\\
                a\\
                b\\
                0
      \end{array}
      \!\!\!
   \right]\!
   :\,
   a,b \in \mathbb{R}
   \right\}
   \;\;\;\;  .
\end{equation}
\smallskip

\noindent One can conclude from here that the projection operator $\hat{P}_{z}^{\mathrm{Diff}}$  associated with the proposition $\mathrm{Diff}_z$ can be presented as the sum:\smallskip

\begin{equation} \label{32} 
   \hat{P}_{z}^{\mathrm{Diff}}
   =
   \hat{P}_{z}^{\uparrow\downarrow} + \hat{P}_{z}^{\downarrow\uparrow}
   \;\;\;\;  .
\end{equation}
\smallskip

\noindent Likewise, one can get\smallskip

\begin{equation} \label{33} 
   |0,0_x\rangle
   =
   \left(\!
   -\frac{1}{\sqrt{2}}
   \right)\!\!
      \left[
         \!\!\!
         \begin{array}{l}
                     0\\
                     1\\
            \bar{1}\\
                     0
         \end{array}
         \!\!\!
      \right]
   \in
   \mathrm{ran}(\hat{P}_{x}^{\mathrm{Diff}})
   =
   \left\{
   \left[
      \!\!\!
      \begin{array}{r}
                a-c\\
                   b\\
                  -b\\
               a+c
      \end{array}
      \!\!\!
   \right]\!
   :\,
   a,b,c \in \mathbb{R}
   \right\}
   \;\;\;\;  ,
\end{equation}
\smallskip

\noindent where\smallskip

\begin{equation} \label{34} 
   \hat{P}_{x}^{\mathrm{Diff}}
   =
   \hat{P}_{x}^{\uparrow\downarrow} + \hat{P}_{x}^{\downarrow\uparrow}
   =
   \frac{1}{4}\!
      \left[
         \!\!\!
         \begin{array}{c c c c}
                     1 & \bar{1} &          1 &          1\\
            \bar{1} &          1 & \bar{1} &          1\\
                     1 & \bar{1} &          1 & \bar{1}\\
            \bar{1} &          1 & \bar{1} &          1
         \end{array}
         \!\!\!
      \right]
   +
   \frac{1}{4}\!
      \left[
         \!\!\!
         \begin{array}{c c c c}
                     1 &          1 & \bar{1} & \bar{1}\\
                     1 &          1 & \bar{1} & \bar{1}\\
            \bar{1} & \bar{1} &          1 &          1\\
            \bar{1} & \bar{1} &          1 &          1
         \end{array}
         \!\!\!
      \right]
   =
   \frac{1}{2}\!
      \left[
         \!\!\!
         \begin{array}{c c c c}
                     1 &          0 &          0 & \bar{1}\\
                     0 &          1 & \bar{1} &          0\\
                     0 & \bar{1} &          1 &          0\\
            \bar{1} &          0 &          0 &          1
         \end{array}
         \!\!\!
      \right]
   \;\;\;\;  ,
\end{equation}

\begin{equation} \label{35} 
   \mathrm{ran}(\hat{P}_{x}^{\uparrow\downarrow})
   =
   \left\{
   \left[
      \!\!\!
      \begin{array}{r}
         a\\
        -a\\
         a\\
        -a
      \end{array}
      \!\!\!
   \right]
   :\,
   a \in \mathbb{R}
   \right\}
   \;
   ,
   \;
   \mathrm{ran}(\hat{P}_{x}^{\downarrow\uparrow})
   =
   \left\{
   \left[
      \!\!\!
      \begin{array}{r}
         a\\
         a\\
        -a\\
        -a
      \end{array}
      \!\!\!
   \right]
   :\,
   a \in \mathbb{R}
   \right\}
   \;\;\;\;  ,
\end{equation}
\smallskip

\noindent and\smallskip

\begin{equation} \label{36} 
   |0,0_y\rangle
   =
   \left(\!
   -\frac{i}{\sqrt{2}}
   \right)\!\!
      \left[
         \!\!\!
         \begin{array}{l}
                     0\\
                     1\\
            \bar{1}\\
                     0
         \end{array}
         \!\!\!
      \right]
   \in
   \mathrm{ran}(\hat{P}_{y}^{\mathrm{Diff}})
   =
   \left\{
   \left[
      \!\!\!
      \begin{array}{r}
                   a\\
                   b\\
                  -b\\
                   a
      \end{array}
      \!\!\!
   \right]\!
   :\,
   a,b \in \mathbb{C}
   \right\}
   \;\;\;\;  ,
\end{equation}
\smallskip

\noindent in which\smallskip

\begin{equation} \label{37} 
   \hat{P}_{y}^{\mathrm{Diff}}
   =
   \hat{P}_{y}^{\uparrow\downarrow} + \hat{P}_{y}^{\downarrow\uparrow}
   =
   \frac{1}{4}\!
      \left[
         \!\!\!
         \begin{array}{r r r r}
                    1 &             i &             -i &          1\\
                    -i &            1 &  \bar{1} &          -i\\
                     i & \bar{1} &             1 &           i\\
                    1 &             i &             -i &          1
         \end{array}
         \!\!\!
      \right]
   +
   \frac{1}{4}\!
      \left[
         \!\!\!
         \begin{array}{r r r r}
                     1 &            -i &             i &           1\\
                      i &            1 & \bar{1} &            i\\
                     -i & \bar{1} &            1 &          -i\\
                     1 &            -i &             i &           1
         \end{array}
         \!\!\!
      \right]
   =
   \frac{1}{2}\!
      \left[
         \!\!\!
         \begin{array}{c c c c}
                     1 &          0 &          0 & \bar{1}\\
                     0 &          1 & \bar{1} &          0\\
                     0 & \bar{1} &          1 &          0\\
            \bar{1} &          0 &          0 &          1
         \end{array}
         \!\!\!
      \right]
   \;\;\;\;  ,
\end{equation}

\begin{equation} \label{38} 
   \mathrm{ran}(\hat{P}_{y}^{\uparrow\downarrow})
   =
   \left\{
   \left[
      \!\!\!
      \begin{array}{r}
         a\\
       -ia\\
        ia\\
         a
      \end{array}
      \!\!\!
   \right]
   :\,
   a \in \mathbb{C}
   \right\}
   \;
   ,
   \;
   \mathrm{ran}(\hat{P}_{y}^{\downarrow\uparrow})
   =
   \left\{
   \left[
      \!\!\!
      \begin{array}{r}
         a\\
        ia\\
       -ia\\
         a
      \end{array}
      \!\!\!
   \right]
   :\,
   a \in \mathbb{C}
   \right\}
   \;\;\;\;  .
\end{equation}
\smallskip

\noindent Consequently, one gets\smallskip

\begin{equation} \label{39} 
   |0,0_j\rangle
   =
   C\!
      \left[
         \!\!\!
         \begin{array}{l}
                     0\\
                     1\\
            \bar{1}\\
                     0
         \end{array}
         \!\!\!
      \right]
   \in
   \left\{
   \left[
      \!\!\!
      \begin{array}{r}
                0\\
                a\\
                b\\
                0
      \end{array}
      \!\!\!
   \right]\!
   :\,
   a,b \in \mathbb{C}
   \right\}
   \subseteq
   \left\{
   \left[
      \!\!\!
      \begin{array}{r}
                a-c\\
                   b\\
                  -b\\
               a+c
      \end{array}
      \!\!\!
   \right]\!
   :\,
   a,b,c \in \mathbb{C}
   \right\}
   \subseteq
   \left\{
   \left[
      \!\!\!
      \begin{array}{r}
                   a\\
                   b\\
                  -b\\
                   a
      \end{array}
      \!\!\!
   \right]\!
   :\,
   a,b \in \mathbb{C}
   \right\}
   \;\;  ,
\end{equation}

\begin{equation} \label{40} 
   |0,0_j\rangle
   =
   C\!
      \left[
         \!\!\!
         \begin{array}{l}
                     0\\
                     1\\
            \bar{1}\\
                     0
         \end{array}
         \!\!\!
      \right]
   \notin
   \left\{  
      \begin{array}{r}
         \left\{       
         \left[
         \!\!\!
            \begin{array}{r}
                      0\\
                      a,0\\
                     0,a\\
                      0
            \end{array}
         \!\!\!
      \right]\!
      :\,
      a \in \mathbb{C}
      \right\}\\
         \left\{       
         \left[
         \!\!\!
            \begin{array}{r}
                      a\\
                      \mp a\\
                     \pm a\\
                      -a
            \end{array}
         \!\!\!
      \right]\!
      :\,
      a \in \mathbb{C}
      \right\}\\
         \left\{       
         \left[
         \!\!\!
            \begin{array}{r}
                      a\\
                      \mp ia\\
                     \pm ia\\
                      a
            \end{array}
         \!\!\!
      \right]\!
      :\,
      a \in \mathbb{C}
      \right\}
      \end{array}
   \right. 
   \;\;\;\;  ,
\end{equation}
\smallskip

\noindent where $C \in \mathbb{C}$. Under the supervaluationist postulation (\ref{20}), this brings the following valuations in the singlet state $ |0,0_j\rangle$:\smallskip

\begin{equation} \label{41} 
   |0,0_j\rangle
   \in
   \mathrm{ran}(\hat{P}_{j}^{\mathrm{Diff}})
   \;
   \iff
   \;
   v(\hat{P}_{j}^{\uparrow\downarrow} + \hat{P}_{j}^{\downarrow\uparrow})
   =
   {[\![ \, \mathrm{Diff}_j \, ]\!]}_v
   =
   1
   \;\;\;\;  ,
\end{equation}

\begin{equation} \label{42} 
   |0,0_j\rangle
   \notin
   \left\{  
      \begin{array}{r}
            \!\!\!
                      \mathrm{ran}(\hat{P}_{j}^{\uparrow\downarrow})\\
                      \mathrm{ker}(\hat{P}_{j}^{\uparrow\downarrow})\\
                      \mathrm{ran}(\hat{P}_{j}^{\downarrow\uparrow})\\
                      \mathrm{ker}(\hat{P}_{j}^{\downarrow\uparrow})
            \end{array}
   \right. 
   \;
   \iff
   \;
   \begin{array}{r}
           \left\{v(\hat{P}_{j}^{\uparrow\downarrow})\right\} = \left\{  {[\![ \,  \uparrow^{(A)}_j \!\!\!\wedge\! \downarrow^{(B)}_j ]\!]}_v\right\} = \varnothing \\
           \left\{v(\hat{P}_{j}^{\downarrow\uparrow})\right\} = \left\{  {[\![ \,  \downarrow^{(A)}_j\! \!\!\wedge\! \uparrow^{(B)}_j ]\!]}_v\right\} = \varnothing
   \end{array}
   \;\;\;\;  .
\end{equation}
\smallskip

\noindent As follows, even though the proposition $\mathrm{Diff}_j$ has the preexisting value of truth in the state $|0,0_j\rangle$, the statement ``$m^{(A)}_j\!\!=\pm\textonehalf$ and $m^{(B)}_j\!\!=\mp\textonehalf$'' does not have the truth-value at all prior to its verification. Otherwise stated, in the singlet state the valuation ${[\![ \, \mathrm{Diff}_j \, ]\!]}_v =  {[\![ \, \uparrow^{(A)}_j \!\!\!\wedge\! \downarrow^{(B)}_j \underline{\lor}\,\, \downarrow^{(A)}_j\! \!\!\wedge\! \uparrow^{(B)}_j ]\!]}_v$ does not accept \textit{the principle of truth-functionality}, namely, ${[\![ \, \mathrm{Diff}_j \, ]\!]}_v$ cannot be presented as a function of ${[\![ \, \updownarrow^{(A)}_j ]\!]}_v$ and ${[\![ \, \updownarrow^{(B)}_j ]\!]}_v$.\\

\noindent Suppose Alice verifies experimentally that the proposition $\uparrow^{(A)}_z$ is true but the proposition $\downarrow^{(A)}_z$ is false. After Alice’s verification, the spin state of the two-particle system $|\Psi^{(AB)}\rangle$ turns into separable $|\Psi^{(A)}_z\rangle \otimes |\Psi^{(B)}_z\rangle$ where\smallskip

\begin{equation} \label{43} 
   |\Psi^{(A)}_z\rangle
   =
   \left[
         \!\!\!
         \begin{array}{c}
                     1\\
                     0
         \end{array}
         \!\!\!
   \right]
   \in
   \left\{
      \begin{array}{r}
              \mathrm{ran}(|\!\uparrow_{z}^{(A)}\rangle \langle\uparrow_{z}^{(A)}\!\!|) \; \iff \; v(|\!\uparrow_{z}^{(A)}\rangle \langle\uparrow_{z}^{(A)}\!\!|) = {[\![ \, \uparrow^{(A)}_z ]\!]}_v = 1 \\
             \mathrm{ker}(|\!\downarrow_{z}^{(A)}\rangle \langle\downarrow_{z}^{(A)}\!\!|) \; \iff \; v(|\!\downarrow_{z}^{(A)}\rangle \langle\downarrow_{z}^{(A)}\!\!|) = {[\![ \, \downarrow^{(A)}_z ]\!]}_v = 0
      \end{array}
   \right.
   \;\;\;\;   
\end{equation}
\smallskip

\noindent and $|\Psi^{(B)}_z\rangle = \bigl[\begin{smallmatrix} b\\ a \end{smallmatrix} \bigr]$ where $a$ and $b$ are unknown real numbers. This gives\smallskip

\begin{equation} \label{44} 
   |\Psi^{(A)}_z\rangle \otimes |\Psi^{(B)}_z\rangle
   =
   \left[
         \!\!\!
         \begin{array}{c}
                     1\\
                     0
         \end{array}
         \!\!\!
   \right]
   \!\otimes\!
   \left[
         \!\!\!
         \begin{array}{c}
                     b\\
                     a
         \end{array}
         \!\!\!
   \right]
   =
   \left[
      \!\!\!
      \begin{array}{c}
                  b\\
                  a\\
                  0\\
                  0
      \end{array}
      \!\!\!
   \right]
   \in
   \left\{\!\!
      \begin{array}{r}
                  \mathrm{ran}(\hat{P}^{\uparrow\uparrow}_z)\; , \;\; \text{if} \; a=0\\
                  \mathrm{ran}(\hat{P}^{\uparrow\downarrow}_z)\;, \;\; \text{if} \; b=0
      \end{array}
   \right.
   \;\;\;\;  .
\end{equation}
\smallskip

\noindent But since system's state is prepared in the singlet state lying in the sum $\mathrm{ran}(\hat{P}^{\uparrow\downarrow}_{z}) + \mathrm{ran}(\hat{P}^{\downarrow\uparrow}_{z})$, one can infer that after Alice's verification the value of the projection operator $\hat{P}^{\uparrow\downarrow}_{z}$  becomes true, namely, $v(\hat{P}^{\uparrow\downarrow}_{z}) = 1$, which implies $b=0$ and so\smallskip

\begin{equation} \label{45} 
   |\Psi^{(B)}_z\rangle
   =
   \left[
         \!\!\!
         \begin{array}{c}
                     0\\
                     a
         \end{array}
         \!\!\!
   \right]
   \; \iff \; 
   \left\{
      \begin{array}{r}
              v(|\!\uparrow_{z}^{(B)}\rangle \langle\uparrow_{z}^{(B)}\!\!|) = {[\![ \, \uparrow^{(B)}_z ]\!]}_v = 0 \\
              v(|\!\downarrow_{z}^{(B)}\rangle \langle\downarrow_{z}^{(B)}\!\!|) = {[\![ \, \downarrow^{(B)}_z ]\!]}_v = 1
      \end{array}
   \right.
   \;\;\;\;   
\end{equation}
\smallskip

\noindent even without Bob's verification.\\

\noindent It is straightforward that\smallskip

\begin{equation} \label{46} 
   \left\{
   \left[
      \!\!\!
      \begin{array}{r}
                0\\
                a
      \end{array}
      \!\!\!
   \right]\!
   :\,
   a \in \mathbb{R}
   \right\}
   \nsubseteq
   \left\{  
      \begin{array}{l}
         \mathrm{ran}(|\!\uparrow_{x}^{(B)}\rangle \langle\uparrow_{x}^{(B)}\!\!|)
         =
         \left\{       
         \left[
         \!\!\!
            \begin{array}{r}
                      a\\
                      a
            \end{array}
         \!\!\!
        \right]\!
         :\,
         a \in \mathbb{R}
         \right\}\\
         \mathrm{ran}(|\!\downarrow_{x}^{(B)}\rangle \langle\downarrow_{x}^{(B)}\!\!|)
         =
         \left\{       
         \left[
         \!\!\!
            \begin{array}{r}
                      a\\
                     -a
            \end{array}
         \!\!\!
         \right]\!
         :\,
         a \in \mathbb{R}
         \right\}
      \end{array}
   \right. 
   \;\;\;\;  ;
\end{equation}
\smallskip

\noindent so, under the supervaluationist postulation, in the state $|\Psi^{(B)}_z\rangle$ neither $\uparrow_{x}^{(B)}$  nor $\downarrow_{x}^{(B)}$  can carry the truth value, explicitly,\smallskip 

\begin{equation} \label{47} 
   |\Psi^{(B)}_z\rangle
   \notin
   \left\{
      \begin{array}{l}
         \mathrm{ran}(|\!\uparrow_{x}^{(B)}\rangle \langle\uparrow_{x}^{(B)}\!\!|) \\
         \mathrm{ran}(|\!\downarrow_{x}^{(B)}\rangle \langle\downarrow_{x}^{(B)}\!\!|)
      \end{array}
   \right.
   \iff
   \;
   \left\{
      \begin{array}{r}
              \left\{v(|\!\uparrow_{x}^{(B)}\rangle \langle\uparrow_{x}^{(B)}\!\!|)\right\} = \left\{{[\![ \, \uparrow^{(B)}_x  ]\!]}_v\right\} = \varnothing \\
              \left\{v(|\!\downarrow_{x}^{(B)}\rangle \langle\downarrow_{x}^{(B)}\!\!|)\right\} = \left\{{[\![ \, \downarrow^{(B)}_x  ]\!]}_v\right\} = \varnothing
      \end{array}
   \right.
   \;\;\;\;  .
\end{equation}
\smallskip

\noindent Consequently, after the verification of $\uparrow^{(A)}_z$ the statistical population describing the product ${[\![ \, \downarrow^{(B)}_z ]\!]}_v \cdot {[\![ \, \uparrow^{(B)}_x ]\!]}_v$ would have no elements at all:\smallskip

\begin{equation} \label{48} 
   \{{[\![ \, \downarrow^{(B)}_z  ]\!]}_v\} \!\times \{{[\![ \, \uparrow^{(B)}_x  ]\!]}_v\}
   =
   \{1\} \!\times \varnothing
   =
   \varnothing
   \;\;\;\;  .
\end{equation}
\smallskip

\noindent This implies that the statement ``$m^{(B)}_z\!\!=-\textonehalf$ and $m^{(B)}_x\!\!=+\textonehalf$'' would have no truth value, and so the verification (refutation) of $\uparrow^{(A)}_x$ in the next experimental run would not bring the definiteness to the product ${[\![ \, \downarrow^{(B)}_z ]\!]}_v \cdot {[\![ \, \uparrow^{(B)}_x ]\!]}_v$.\\

\section{Conclusion remarks}  

\noindent As the expression (\ref{9}) demonstrates, it is \textit{counter-factual definiteness} that, together with the principle of truth-functionality, led to the bivaluation of the statement ``$m^{(B)}_z\!\!=\mp\textonehalf$ and $m^{(B)}_x\!\!=\pm\textonehalf$''.\\

\noindent Then again, from the point of view of the logical matrix which fixes a model of logic \cite{Siegfried}, counter-factual definiteness can be interpreted as the assertion that the experimentally testable propositions -- like $\updownarrow^{(B)}_z$ and $\updownarrow^{(B)}_x$ -- possess intrinsic truth values that exist even when these propositions have not been verified (see to that end the definition of counter-factual definiteness in \cite{Pearl, Helland, Zukovski}).\\

\noindent This suggests that to permit realist interpretations of quantum mechanics (whose characteristical details can be found, e.g., in papers \cite{Mermin, Kellner}), models of logic underpinning such interpretations must reject preexisting truth-values in general -- as the described in this paper quantum supervaluationist EPR account does.\\

\bibliographystyle{References}
\bibliography{QS_account_of_EPR_ref}

\end{document}